\documentclass[final,3p,times]{elsarticle}

\usepackage[hidelinks]{hyperref}
\usepackage{lineno}

\usepackage{algpseudocode}
\usepackage{layouts}
\usepackage{mathtools,bm}
\usepackage[usestackEOL]{stackengine}

\modulolinenumbers[5]
\usepackage{multirow} 
\usepackage{pgf}
\usepackage{lineno}
\usepackage[para]{threeparttable}
\modulolinenumbers[5]
\journal{Manuscript accepted for publication in Computer Physics Communications}

\usepackage{amsmath}
\usepackage{placeins}
\usepackage{xcolor}
\usepackage{algorithm}
\usepackage{tikz}
\usepackage{nccmath}
\usetikzlibrary{chains}

\tikzset{
every picture/.style={
  execute at end picture={
    \path (current bounding box.south west) +(-3pt,-3pt) (current bounding box.north east) +(3pt,3pt);
    }
  }
}

\definecolor{tikzlightgray}{HTML}{e6e6e6}

\usepackage[normalem]{ulem}
\usepackage{amssymb}
\begin{document}

\begin{frontmatter}

\title{A pencil-distributed finite-difference solver for extreme-scale calculations of turbulent wall flows at high Reynolds number}

\author[mymainaddress]{Rafael Diez\corref{mycorrespondingauthor}}
\cortext[mycorrespondingauthor]{Corresponding author}
\ead{R.G.DiezSanhueza-1@tudelft.nl}
\author[mymainaddress]{Jurriaan Peeters}
\author[mymainaddress]{Pedro Costa}

\address[mymainaddress]{Process \& Energy Department, Delft University of Technology, \\Leeghwaterstraat 39, 2628 CB Delft, The Netherlands}

\begin{abstract}

We present a computational method for extreme-scale simulations of incompressible turbulent wall flows at high Reynolds numbers. The numerical algorithm extends a popular method for solving second-order finite differences Poisson/Helmholtz equations using a pencil-distributed parallel tridiagonal solver to improve computational performance at scale. The benefits of this approach were investigated for high-Reynolds-number turbulent channel flow simulations, with up to about 80 billion grid points and 1024 {GPUs} on the European flagship supercomputers Leonardo and LUMI. An additional GPU porting effort of the entire solver had to be undertaken for the latter. Our results confirm that, while 1D domain decompositions are favorable for smaller systems, they become inefficient or even impossible at large scales. This restriction is relaxed by adopting a pencil-distributed approach. The results show that, at scale, the revised Poisson solver is about twice as fast as the baseline approach with the full-transpose algorithm for 2D domain decompositions. Strong and weak scalability tests show that the performance gains are due to the lower communication footprint. Additionally, to secure high performance when solving for wall-normal implicit diffusion, we propose a reworked flavor of parallel cyclic reduction (PCR) that is split into pre-processing and runtime steps. During pre-processing, small sub-arrays with independent 1D coefficients are computed by parallel GPU threads, without any global GPU communication. Then, at runtime, the reworked PCR enables a fast solution of implicit 1D diffusion without computational overhead. Our results show that the entire numerical solver, coupled with the PCR algorithm, enables extreme-scale simulations with 2D pencil decompositions, which do not suffer performance losses even when compared to the best 1D slab configurations available for smaller systems.
\end{abstract}


\end{frontmatter}

\section{Introduction}

Turbulent flows at high Reynolds numbers are among the most complex and prevalent problems in engineering and physics. While numerous flows at low Reynolds numbers may be studied using simple analytical or numerical models, many flows found in nature and industry operate at high Reynolds numbers in a turbulent regime. These flows exhibit complex behavior, which is difficult to predict using existing correlations or upscaled models. Fundamental understanding of turbulence that can improve engineering models requires direct numerical simulations (DNS), where the chaotic and multi-scale flow dynamics are fully resolved up to the smallest temporal and spatial scales. While this has a large computational cost, the exponential growth in computing power during the last decades, along with the development of efficient numerical methods, have enabled the simulation of flows at increasingly high Reynolds numbers. Indeed, following the developments since the first DNS of isotropic turbulence by \citet{Orszag72} in $1972$, it is now possible to perform large-scale simulations with trillions of grid points in modern supercomputers \cite{Yeung_trillion1,Ishihara_trillion2}.

We are experiencing yet another breakthrough, thanks to the recent proliferation of general-purpose GPU-based supercomputers \cite{hpc1_Reed,hpc2_Nickolls}. GPUs are known to perform well in tasks that only require simple arithmetic operations or RAM access patterns \cite{cuda_fortran_book}, which are common in computational fluid dynamics (CFD). They have much higher throughput than CPUs by allowing many parallel threads to perform the same operation per clock cycle. Thus, when successfully ported, GPU-accelerated solvers can easily outperform multi-core CPU solvers \cite{salvadore13,kim23}, {enabling} the numerical solution to complex problems at much lower costs. However, large-scale CFD problems need to operate at scale, in a distributed-memory paradigm where the data is distributed among many GPUs. This introduces new challenges, as many-GPU systems are more prone to feature performance bottlenecks associated with intra and internode communication, which may require adjustments in the numerical algorithm. Let us consider the current European pre-exascale flagship supercomputers Leonardo and LUMI. {A summary of the GPU specifications for both supercomputers can be found in Table~\ref{table_GPU_nodes}. In the case of LUMI, it is important to highlight that each MI250x is split into two Graphics Compute Dies (GCDs)\footnote{{In this work, we consider each GCD in LUMI to be an independent GPU (unless explicitly noted), since they are exposed to the user's software as separate devices.}} \cite{amd_mi250x}, each with roughly similar characteristics as a NVIDIA A100 GPU in terms of memory and processing power. The internal memory bandwidths for both AMD and NVIDIA GPUs are much faster than intra- and inter-node communication. Therefore, algorithms that minimize device-to-device communication can be optimal, even if they slightly increase the number of arithmetic operations or HBM (High Memory Bandwidth) usage per device. Additionally, opportunities for code optimization can be found by selecting algorithms that replace inter-node communication with faster intra-node data transfers.}

\FloatBarrier
\begin{table}[!ht]
\centering
\begin{threeparttable}
\caption{{Specifications for the GPU nodes in the Leonardo and LUMI supercomputers \cite{nvidia_A100,amd_mi250x,a2a_lumi_leo_bandwidth,amd_data_mov}. The data formally corresponds to the Booster partition in Leonardo, and the LUMI-G nodes for LUMI. The abbreviation ``Gb'' refers to Gigabit, whereas ``GB''/``TB'' corresponds to Gigabyte/Terabyte.}}
\label{table_GPU_nodes}
\def\arraystretch{1.2}
\begin{tabular}{c c c}\\ 
\hline
 & Leonardo & LUMI \\
\hline
GPU model & NVIDIA A100 & AMD MI250X \\
Number of devices & 4 GPUs & 8 GCDs (2 per GPU) \\
Inter-node bandwidth & 100 Gb/s & 100 Gb/s\tnote{$\dagger$} \\
Intra-node bandwidth & 800 Gb/s & 400 – 1600 Gb/s\tnote{$\dagger$} \\
Internal memory bandwidth & 1.6 TB/s & 1.6 TB/s\tnote{$\dagger$} \\
HBM & 64 GB & 64 GB\tnote{$\dagger$} \\
\multirow{2}{*}{FP64 Peak Performance} & 9.7 TFLOPs & \multirow{2}{*}{23.95 TFLOPs\tnote{$\dagger$}} \\
 & 19.5 TFLOPs with Tensor Cores & \\
\hline
\end{tabular}
     \begin{tablenotes}
       \item [$\dagger$] Data for each GCD in LUMI. \\
     \end{tablenotes}
\end{threeparttable}
\end{table}
\FloatBarrier

Several works have performed large-scale turbulent flow simulations using multi-GPU configurations. Compressible flow solvers, for instance, contain many fully explicit calculations that can be readily parallelized using GPUs. The DNS solver \emph{STREAmS} \cite{streams} can use multi-GPU systems to simulate compressible wall-bounded flows, taking into account complex effects such as shock-wave interactions. In \emph{URANOS} \cite{uranos}, a compressible flow solver is developed for large-scale simulations using various modelling frameworks, and several possible choices for the discretization schemes. In incompressible flow solvers, GPU porting for distributed-memory calculations at scale faces an extra challenge. Typically, a major performance bottleneck is solving a large linear system associated with the pressure Poisson equation {to ensure incompressibility}. Nevertheless, several recent works have shown great progress in the development of multi-GPU solvers. Focusing on spectral or finite-difference approaches, an example is the AFiD-GPU code \cite{afid_gpu} for large-scale simulations of wall-bounded flows using multi-GPU (or multi-CPU) configurations. Another example is the CaNS code, which is used in the present study \cite{cpu_cans_18,gpu_cans_21}. CaNS is an incompressible DNS solver, which is compatible with various types of boundary conditions for canonical flow cases in rectangular grids, such as isotropic turbulence or wall-bounded flows. This solver is compatible with both multi-GPU and multi-core CPU architectures, and has been recently re-ported to GPUs porting using OpenACC and the hardware-adaptive \emph{cuDecomp} library for GPU communications at scale \cite{cudecomp}. This library allows pencil-distributed solvers that require collective transpose operations to perform runtime autotuning and determine the optimal domain decomposition and GPU communication backend. 
{
A simple FFT-based finite-differences numerical solver like CaNS requires two types of communication operations: halo exchanges and transposes. Halo exchanges are relatively simple, standard operations, where each task exchanges boundary values with its neighbors.  
Transposes are more expensive \emph{all-to-all} collective operations, where 3D data of a field is redistributed among MPI tasks such that all cells aligned in a specific dimension are local to a single process/task. This is important while performing, for instance, fast Fourier-based transformations in spectral Poisson/Helmholtz solvers, since FFT algorithms require frequent access to the arrays being transformed. Naturally, all transpose operations involving 3D arrays are expensive, and often the major performance bottleneck. 

Second-order FFT-based finite-difference solvers such as those used in AFiD and CaNS require performing FFTs along two directions, and the solution of a resulting tridiagonal system along the other direction, which is typically the wall-normal one in the case of wall-bounded flows with one inhomogeneous direction. The solution of the tridiagonal system has been typically performed using transpose operations, such that the whole system is local to each task and serially solved. However, there is possible room for improvement here by exploiting a parallel tridiagonal solver that avoids this collective operation.} Notably, an interesting approach was presented by \citet{lazlo} and exploited in \cite{YANG23_PASCAL_TDMA_v20,pascal_tdma,gong2022high,YANG23_DTDMA_microclimate}, which showed compelling performance gains at scale. In short, this method uses a hybrid Thomas--parallel cyclic reduction (PCR) algorithm that effectively converts the tridiagonal system to be solved in parallel into a series of smaller systems that can be solved independently, coupled to a smaller problem to be solved collectively for the first and last unknowns of each small system \cite{gong2022high}. 

\begin{figure*}[h]
\centering
\includegraphics[width=104mm]{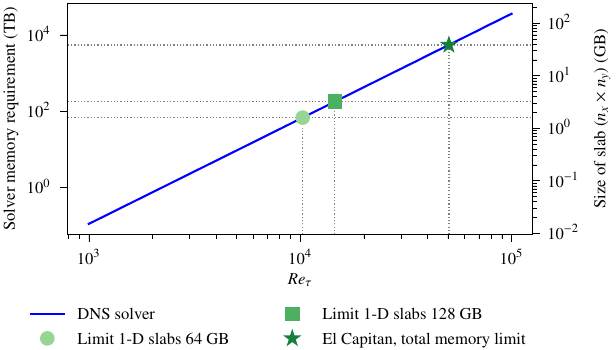}
\caption{Estimates of the total memory requirements for double-precision DNS runs of turbulent channel flows on a $L_x~\times~L_y~\times~L_z~=~12.8~\times~6.4~\times~2$ domain at different friction Reynolds numbers ($Re_{\tau}$), as well as the size of a single slice $n_x \times n_y$ of the DNS domain in HBM for all points in the streamwise ($x$) and spanwise ($y$) directions. The calculations are performed using the domain partition algorithms of the DNS solver CaNS \cite{gpu_cans_21} and the parallel decomposition library \emph{cuDecomp} \cite{cudecomp}. The symbols denote the limits where a 1D slab decomposition becomes impossible for GPUs with {64 GB and 128 GB} of memory, respectively, which is typical of current high-end HPC GPUs, along with the {total GPU memory of the largest supercomputer as of 2025: El Capitan \cite{top500}. This marks an upper bound of the maximum Reynolds number that could be investigated with current computational resources: $\mathrm{Re}_\tau \approx 50,000$.}}
\label{fig_ram_usage}
\end{figure*}

Most works exploring PCR in this context have adopted a 1D parallelization \cite{pascal_tdma,YANG23_DTDMA_microclimate}, with slabs parallel to directions of FFT-based synthesis. This is efficient and was proven to work up to a certain scale. However, as the flow Reynolds number increases, it becomes impossible to resort to a 1D parallelization. As an example, Figure~\ref{fig_ram_usage} presents the total memory requirement in a DNS solver (CaNS), to simulate turbulent channel flows at increasing friction Reynolds number ($Re_{\tau}$), as well as the total size of a single $n_x \times n_y$ wall-parallel slice. Expectedly, as $Re_{\tau}$ increases, in addition to stricter time steps restrictions, the number of grid cells increases roughly as $N_{x,y}\propto Re_{\tau}$ in the streamwise ($x$) and spanwise ($y$) directions of the channel flow, and $N_{z}\propto Re_{\tau}^{3/4}$ in the wall-normal direction ($z$) \cite{PIROZZOLI21,pope2001turbulent}. Hence, as the Reynolds number increases, the thickness of a wall-parallel slab that fits a fixed amount of grid points (e.g., dictated by the RAM constrains of a GPU or CPU device) becomes ever thinner, until it becomes impossible to decompose the domain further. This is particularly problematic in wall-bounded turbulence, where the number of grid points along the wall-parallel directions should be larger than in the wall-normal one \cite{PIROZZOLI21}. Yet, the same is bound to happen in other turbulent flows (e.g., homogeneous isotropic turbulence) at sufficiently high Reynolds number.

{Consequently}, even with the ever-increasing memory capacity of GPUs, for DNS of high Reynolds number flows with this type of approach, one may be bound to adopt a 2D pencil-like domain decomposition. Leveraging a less communication-intensive approach for solving the Poisson equation, while retaining a pencil-distributed decomposition, is precisely the motivation of the present work. {One alternative within this context is to replace serial TDMA (tridiagonal matrix algorithms) by parallel methods, with a lower communication footprint.} We develop such an approach based on a PCR-TDMA method {(named P-TDMA hereafter)}, and test it on the CaNS solver with a focus on many-GPU calculations at scale. To run our solver on an AMD-based system like LUMI, an additional porting effort was undertaken, which we will also describe here. We report the revised solver's performance and scalability for large-scale DNS of turbulent channel flow at high Reynolds number. Moreover, we show that particular care should be taken for wall-normal implicit diffusion, to secure high performance at scale. Our results show an almost $2\times$ speedup at scale for the Poisson solver with 2D decomposition, which significantly improves the overall solver performance on large-scale GPU-based supercomputers.

This manuscript is organized as follows. Section~\ref{section_methodology} describes the governing equations, the discretization scheme, and the implementation of the Poisson/Helmholtz solver, including the cross-platform effort to run on AMD-based systems. Then Section~\ref{chapter_results} discusses the study and the scalability benchmarks. Finally, Section~\ref{section_conclusions} presents the conclusions.

\FloatBarrier
\section{Methodology}
\label{section_methodology}
\subsection{Governing equations and numerical discretization}
\label{section_discrete_eqs}
The current numerical framework solves the incompressible Navier-Stokes equations,
\begin{equation}
    \nabla \cdot \textbf{u} = 0,
    \label{NS_cont}
\end{equation}
\begin{equation}
    \partial_t \textbf{u} + \left( \textbf{u} \cdot \nabla \right) \textbf{u} = - \nabla p + \nu \nabla^2 \textbf{u},
    \label{NS_momen}
\end{equation}
where $\textbf{u}$ and $p$ correspond to the fluid velocity vector and the pressure scaled by the fluid density; $\nu$ is the fluid kinematic viscosity. The numerical scheme is based on an incremental pressure correction scheme {or fractional-step method} \cite{chorin67,KIM85}, where a prediction velocity $\mathbf{u}^*$ is first calculated by integrating the momentum equation in time, and continuity is imposed using a correction pressure $\Phi$, which is obtained from the solution of a Poisson equation. The equations were solved on a rectangular box using a structured Cartesian grid with a staggered (MAC) arrangement for the velocity and pressure grid cells. As per the restriction of the Poisson solver, we employ uniform spacing along two Cartesian directions ($x$,$y$) and non-uniform spacing in the third spatial direction ($z$). Second-order finite differences are used for spatial discretization. This has several advantages with respect to higher-order methods, such as being computationally efficient while still enabling simulations with similar fidelity as spectral discretization methods in practice \cite{moin_verzicco}, and being flexible and easily extended with numerical techniques for handling complex geometries like the immersed-boundary method \cite{fadlun00,BreugemBoersma,uhlmann05}, or multi-fluid flows \cite{Tryggvason-et-al-2011}. Finally, Wray's low-storage Runge-Kutta scheme \cite{Wray-NASA-1986} is used for temporal discretization. The numerical scheme is presented below in semi-discrete form:

\begin{equation}
    \textbf{u}^{*} =  \textbf{u}^k + \Delta t \left( \alpha_k \left( \mathcal{A} \textbf{u}^k + \nu \mathcal{L} \textbf{u}^k \right) + \beta_k \left( \mathcal{A} \textbf{u}^{k-1} + \nu \mathcal{L} \textbf{u}^{k-1} \right) - \gamma_k \mathcal{G} p^{k-1/2} \right),
    \label{proj1_ustar}
\end{equation}

\begin{equation}
    \mathcal{L} \Phi = \frac{\mathcal{D} \textbf{u}^{*}}{\gamma_k \Delta t},
    \label{proj1_Lphi}
\end{equation}

\begin{equation}
    \textbf{u}^{k+1} = \textbf{u}^{*} - \gamma_k \Delta t \mathcal{G} \Phi,
    \label{proj1_ukp1}
\end{equation}

\begin{equation}
    p^{k+1/2} = p^{k-1/2} + \Phi.
    \label{proj1_Pkp1}
\end{equation}

where $\alpha$, $\beta$, and $\gamma$ refer to the coefficients of the RK3 scheme, which are given by: $\alpha = \lbrace 8/15,\ 5/12,\ 3/4 \rbrace$, $\beta = \lbrace 0,\ -17/60,\ -5/12 \rbrace$ and $\gamma = \alpha + \beta$; the index $k$ refers to the RK3 sub-iteration index $k=\lbrace0,1,2\rbrace$, and $\Delta t$ is the time step. For flows at very low Reynolds numbers, or highly refined grids, the time step size $\Delta t$ can be prohibitively small if diffusive terms are integrated in time explicitly. In such cases, it may be preferable to perform an implicit discretization of the diffusion terms at the cost of solving an extra Helmholtz equation per velocity component, as illustrated below:

\begin{equation}
    \textbf{u}^{**} =  \textbf{u}^k + \Delta t \left( \alpha_k  \mathcal{A} \textbf{u}^k + \beta_k \mathcal{A} \textbf{u}^{k-1} + \gamma_k \left( - \mathcal{G} p^{k-1/2} + \nu \mathcal{L} \textbf{u}^k \right) \right),
    \label{proj2_ustarstar}
\end{equation}
\begin{equation}
    \textbf{u}^{*} - \gamma_k \frac{\nu \Delta t}{2} \mathcal{L} \textbf{u}^* =  \textbf{u}^{**} - \gamma_k \frac{\nu \Delta t}{2} \mathcal{L} \textbf{u}^k, 
    \label{proj2_ustar}
\end{equation}
\begin{equation}
    \mathcal{L} \Phi = \frac{\mathcal{D} \textbf{u}^{*}}{\gamma_k \Delta t},
    \tag{\ref{proj1_Lphi}}
\end{equation}
\begin{equation}
    \textbf{u}^{k+1} = \textbf{u}^{*} - \gamma_k \Delta t \mathcal{G} \Phi,
    \label{proj2_ukp1}
\end{equation}
\begin{equation}
    p^{k+1/2} = p^{k-1/2} + \Phi - \gamma_k \frac{\nu \Delta t}{2} \mathcal{L} \Phi.
    \label{proj2_Pkp1}
\end{equation}
Note that, indeed, Eq.~\eqref{proj2_ustar} is a Helmholtz equation for the prediction velocity $\mathbf{u}^*$, which is implicit in the three spatial directions (x,y,z). While this equation can be solved efficiently using the fast direct methods presented in Section~\ref{section_poi_helm_solver}, the computational overhead is still considerable. Fortunately, in many cases, the time step constraints for $\Delta t$ are due to fine grid spacing only along the non-uniform grid direction (here, $z$). This is particularly true for wall-bounded flows with one inhomogeneous direction, such as pipes or channels, which require fine grid spacing near the walls \cite{afid_gpu, cpu_cans_18}. In these cases, one can discretize only the wall-normal diffusion term implicitly, and replace Eq.~\eqref{proj2_ustar} with a one-dimensional system per velocity component:
\begin{equation}
     \textbf{u}^{*} - \gamma_k \frac{\nu \Delta t}{2} \mathcal{L}_{z} \textbf{u}^* =  \textbf{u}^{**} - \gamma_k \frac{\nu \Delta t}{2} \mathcal{L}_z \textbf{u}^k, 
    \label{proj3_impl1d}
\end{equation}
where $\mathcal{L}_z$ denotes the discrete Laplacian term associated with the $z$ direction. This is numerically much cheaper, as the second-order finite-difference discretization of this equation requires the solution of a simple tridiagonal system.%

\subsection{Numerical solution of the Poisson/Helmholtz equation}
\label{section_poi_helm_solver}

\subsubsection{Fourier-based synthesis}
The solution of the Poisson equation for the pressure comprises some of the numerical algorithm's most computation and communication-intensive steps. Here, Eqs.~\eqref{proj1_Lphi} and~\eqref{proj2_ustar} are solved using the method of eigenfunctions expansions, which allows for fast, direct solutions by leveraging the FFT algorithm \cite{schumann88,cpu_cans_18}. After performing a Fourier-based synthesis of the Poisson/Helmholtz equation along directions $x$ and $y$, the following system of tridiagonal equations can be obtained along the non-uniform grid direction $z$ for a grid cell with index $i,j,k$:

\begin{equation}
     \left(\lambda_i / \Delta x^2 + \lambda_j / \Delta y^2 \right) \tilde{\Phi}_{i,j,k} + \left( \eta_{k-1} \tilde{{\Phi}}_{i,j,k-1} + \eta_{k} \tilde{{\Phi}}_{i,j,k} + \eta_{k+1} \tilde{{\Phi}}_{i,j,k+1} \right) =  \tilde{f}_{i,j,k},
    \label{fijk_eq}
\end{equation}

where the tilde $(\,\Tilde{\,}\,)$ denotes two successive discrete Fourier-based (i.e., Fourier/cosine/sine) transforms applied to a variable along the $x$ (index $i$) and $y$ (index $j$) directions. Note that each $(i,j)$ pair corresponds to a tridiagonal system along the non-uniform direction ($z$, index $k$). The coefficients $\lambda_i$ and $\lambda_j$ are the second-order accurate eigenvalues (or modified wavenumbers); see, e.g., \cite{schumann88}; $\Delta x$ and $\Delta y$ correspond to the uniform grid spacing in the $x$ and $y$ directions, whereas the set of coefficients $\eta$ represent the finite-different discretization of the $\mathcal{L}_{z}$ operator along $z$. While the method of eigenfunctions expansions aims at exploiting the FFT algorithm, it still allows for multiple combinations of boundary conditions representative of different classes of canonical turbulent flows, from isotropic turbulence to several boundary-free and wall-bounded shear flows.

After obtaining $\tilde{{\Phi}}_{i,j,k}$ from Eq.~\eqref{fijk_eq}, the final solution ($\Phi_{i,j,k}$) is easily computed from the inverse Fourier-based synthesis. The numerical methods and algorithms to solve Eq.~\eqref{fijk_eq} are the key parts of the present work and will be discussed next in Section~\ref{section_method_tripar}. Finally, there are noteworthy nuances in implementing fast real-to-complex/complex-to-real (Fourier) and real-to-real (sine/cosine) transforms on GPUs in a unified framework, which we describe in \ref{appendix_fourier}.

\subsubsection{Original distributed-memory solution} \label{section_method_tripar}

\begin{figure*}[h]
\centering
\includegraphics[width=120mm]{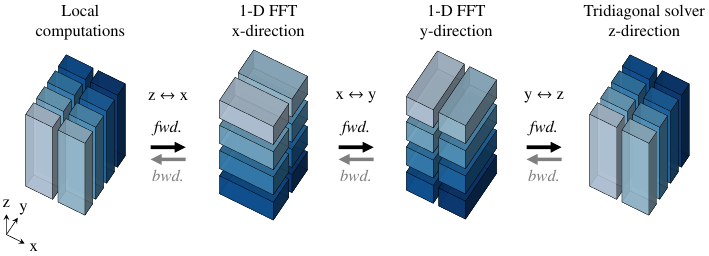}
\caption{Schematic representation of a FFT-based linear solver using a 2D pencil decomposition for the DNS domain. The black arrows indicate the global transpose operations for the data stored among different MPI tasks. The color of each 2D pencil represents a different MPI task. Forward (\emph{fwd.}) operations are first performed from left to right following the direction of the black arrows. Then, all transpose operations are reversed, and inverse/backward (\emph{bwd.}) Fourier-based transforms are performed, as indicated by the gray arrows. {Note that the first transpose operation is often implemented as two consecutive transposes: $z~\to~y$ and $y~\to~x$.}}
\label{fig_FTM_2d_pencils}
\end{figure*}

The numerical solution of the second-order finite-difference Poisson/Helmholtz equations in rectangular grids is solved using FFT-based methods in a distributed-memory setting. The original approach used to solve this problem is described below\footnote{It is important to note that, while the code CaNS allows for an arbitrary default pencil orientation (i.e., outside the pressure solver), we start from $Z$-aligned pencils since this minimizes the number of collective communications when solving the momentum equation with $z$-implicit diffusion. Starting from $x$-aligned pencils would avoid transpose operations in the Poisson solver, but many additional transpose operations would be required for inverting a tridiagonal system per velocity component.}, and illustrated in Figure~\ref{fig_FTM_2d_pencils}.  The following steps are taken:

\begin{enumerate}
\item Compute right-hand side term $d_{i,j,k}$ of eq.~\eqref{fijk_eq} in $z$-aligned pencils.
\item Transpose data to $x$-aligned pencils and perform $N_yN_z$ Fourier-based transforms along $x$.
\item Transpose data to $y$-aligned pencils and perform $N_xN_z$ Fourier-based transforms along $y$.
\item Transpose data to $z$-aligned pencils and solve the resulting $N_xN_y$ tridiagonal systems of equations along $z$.
\item Perform the reciprocate transpose operation as in step~4.
\item Perform the reciprocate inverse transforms and transpose operation as in step~3.
\item Perform the reciprocate {inverse transforms and} transpose operation as in step~2, to obtain the final solution in $z$-aligned pencils.
\end{enumerate}

Here, $N_{x/y/z}$ are the local number of grid cells along $x/y/z$ during the different steps of the algorithm for each MPI task. The transpose operations are an \emph{all-to-all} collective, which may be very expensive. Within this approach, the FFT-based transforms and solution of the tridiagonal systems can be trivially mapped to different parallel (GPU) threads. Note that, whenever a 1D slab-like parallelization is possible, some of the transpose operations shown above would turn into a no-op, making it often desirable. In this regard, {the best-performing slab configurations are those partitioned along $y$. This is convenient, since each GPU can perform Fourier transformations along the $x$-direction, and solve tridiagonal systems of equations along the $z$-direction, without performing additional collective operations. Even with 1D implicit diffusion, only one pair of transposes is required per step: the $x~\leftrightarrow~y$ transposes shown in Figure~\ref{fig_FTM_2d_pencils}.}

\begin{figure*}[h]
\centering
\includegraphics[width=152mm]{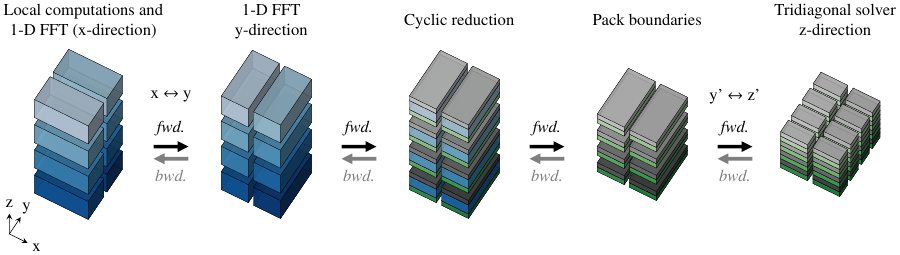}
\caption{Modified parallel tridiagonal solver for large-scale DNS. The first two sub-images follow the same conventions as Figure~\ref{fig_FTM_2d_pencils}, where the color of each partition indicates a different MPI task. In the third and fourth sub-images, the process of cyclic reduction is highlighted, by applying a different color to the boundary values for each partition.}
\label{fig_DTDMA_2d_pencils}
\end{figure*}

While the approach presented in Figure~\ref{fig_FTM_2d_pencils} enables GPU-accelerated DNS of fluid flows on many CPUs/GPUs, the transposing operations may result in a major performance loss. This may be particularly problematic in modern GPU-based systems at scale, {as the inter-node bandwidth is orders of magnitude slower than the GPU memory bandwidth or the intra-node communication}. Hence, we need an approach that: (1) keeps a 2D parallelization, which is unavoidable at scale, and (2) reduces the amount of data used in collective communications to a reasonable minimum. We will explain this approach below.

\subsubsection{Solution with parallel tridiagonal solver} \label{section_poisolve_dtdma_method}

In this approach, we split the computational domain along the $z$ direction, and exploit a parallel tridiagonal solver  as in \cite{lazlo,pascal_tdma,pascal_tcs} to circumvent the large \emph{all-to-all} operations in the previous section. This approach is shown in Figure~\ref{fig_DTDMA_2d_pencils}. This algorithm starts with a cyclic reduction step, such that only information regarding the boundaries of every slice in the $z$-direction must be communicated to other MPI tasks. Subsequently, a tri-diagonal system for the boundary values is constructed. While an \emph{all-to-all} type of collective is still needed, internal data points are not communicated, which can drastically reduce the communication overhead. The steps of this algorithm are summarized as follows:

\begin{enumerate}

\item Compute right-hand side term $d_{i,j,k}$ of eq.~\eqref{fijk_eq} in $x$-aligned pencils.
\item Perform $N_yN_z$ Fourier-based transforms along $x$ (see Figure~\ref{fig_DTDMA_2d_pencils}).

\item Transpose data to $y$-aligned pencils and perform $N_xN_z$ Fourier-based transforms along $y$.

\item Perform $N_xN_y$ cyclic reductions along $z$, pack $2N_xN_y$ boundary values, and transpose the packed boundaries to $z$-aligned pencils.

\item Solve $N_xN_y$ reduced systems of tridiagonal equations along $z$, with local size $2 p_z$.

\item Transpose data to $y$-aligned pencils, unpack $2N_xN_y$ boundary values, and reconstruct the internal solution fields.

\item Perform the reciprocate inverse transforms and transpose operation as in step~3.

\item Perform the reciprocate inverse transforms as in step~2.
\end{enumerate}

In the previous steps, $N_{x/y/z}$ is again the local grid size for each MPI rank in each Cartesian direction, whereas $p_z$ is the number of divisions of the computational domain along the $z$-direction, which would correspond to $p_z=4$ in Figure~\ref{fig_DTDMA_2d_pencils}. The details of the parallel tridiagonal algorithm, and the modifications proposed in this work for the computation of its internal coefficients, are explained in Sections~\ref{section_partridiag}. Clearly, {the} amount of data transferred among MPI tasks is substantially reduced \cite{pascal_tdma}. 
The tridiagonal system of equations found in the right-side of Figure~\ref{fig_DTDMA_2d_pencils} have a size of $2 p_z$, where $p_z$ is the number of partitions of the computational domain along the z-direction. Therefore, the parallel tridiagonal solver should be efficient as long as $2 p_z \ll N_z$, where $N_z$ is the total number of grid points in the $z$-direction.

{Interestingly, in the parallel tridiagonal solver, increasing the number of lateral divisions ($p_y$) favors strong scalability: When $p_y$ is increased, the size of the boundaries (per MPI task) is reduced as $2 n_x n_y / p_y$. Therefore, {doubling} $p_y$ halves the data communicated per task, leading to excellent scalability. In contrast, increasing the number of vertical partitions ($p_z$) does not reduce the MPI workload, and thus it is unfavorable for scalability. This is particularly relevant for 1D slab configurations where $p_y=1$ and $p_z$ is the total of GPUs. Still, 1D slab configurations are optimal when $p_z \ll N_z$.}

\subsubsection{Parallel tridiagonal algorithm} \label{section_partridiag}

Numerous approaches may be considered to parallelize the Thomas algorithm to solve a tridiagonal system (see, e.g., the survey in \cite{pascal_tdma}). Here we adopt the method proposed by \cite{lazlo}, which uses cyclic reduction, combined with a Thomas algorithm for a reduced system. We have illustrated the approach in Figure~\ref{fig_DTDMA_2d_pencils}, and summarize it below.

First, the distributed tridiagonal systems of this form (cf. eq.~\eqref{fijk_eq})
\begin{equation}
     a_{k}~{\phi}_{k-1} + b_{k}~{\phi}_{k} + c_{k}~{\phi}_{k+1} = d_{k},
    \label{eq_tridiag_base}
\end{equation}
are locally reduced to a problem where inner unknowns within the computational subdomain are only a function of the values at its top and bottom boundary:
\begin{equation}
     a'_{k}~{\phi}_{0} + {\phi}_{k} + c'_{k}~{\phi}_{m-1} =  {d'}_{k},
    \label{eq_tridiag_redcyc}
\end{equation}
using a cyclic reduction step. The original set of coefficients and right-hand-side $(a,b,c,d)$ are then reduced to the $(a',c',d')$, where the main diagonal is normalized to have unit weight. Algorithm~\ref{algo_orig_cyclic} describes this approach for completeness, and more details can be found in \cite{lazlo,pascal_tdma,pascal_tcs}.

Second, the sets of values $(a’,c’,d')$ at the boundaries of every domain $k=\{0, m-1\}$ can be grouped and transposed in a collective operation (recall Figure~\ref{fig_DTDMA_2d_pencils}). Then, the standard Thomas algorithm solves the reduced systems of tridiagonal equations for the boundary values of all subdomains. Finally, the boundary data for ${\phi}_{i,j,k}$ can be globally transposed, and the values of $\phi_{i,j,k}$ in the interior of every sub-domain can be reconstructed using eq.~\eqref{eq_tridiag_redcyc}.

\begin{algorithm}
\caption{Cyclic reduction step of the parallel tridiagonal algorithm \cite{lazlo,pascal_tdma}, where $\textbf{d}$ corresponds to the right-hand-side of the system (see Eq.~\eqref{fijk_eq}). %
}
\label{algo_orig_cyclic}
\begin{algorithmic}[1]
\State \textbf{Step 1: Initialization} 
\State\textbf{Input:} $\textbf{a},\textbf{b},\textbf{c},\textbf{d}$
\State $\textbf{a}_0 \leftarrow \textbf{a}_0/\textbf{b}_0$ ;  $\textbf{c}_0 \leftarrow \textbf{c}_0/\textbf{b}_0$ ;  $\textbf{d}_0 \leftarrow \textbf{d}_0/\textbf{b}_0$
\State $\textbf{a}_1 \leftarrow \textbf{a}_1/\textbf{b}_1$ ;  $\textbf{c}_1 \leftarrow \textbf{c}_1/\textbf{b}_1$ ;  $\textbf{d}_1 \leftarrow \textbf{d}_1/\textbf{b}_1$

\For{i=2,...,m-1}
\State $r \leftarrow 1/(\textbf{b}_i - \textbf{a}_i \textbf{c}_{i-1})$
\State $\textbf{d}_i \leftarrow r (\textbf{d}_i - \textbf{a}_i \textbf{d}_{i-1})$
\State $\textbf{c}_i \leftarrow r \textbf{c}_i$
\State $\textbf{a}_i \leftarrow - r \textbf{a}_i \textbf{a}_{i-1}$
\EndFor

\For{i=m-3,...,1}
\State $\textbf{d}_i \leftarrow \textbf{d}_i - \textbf{c}_i \textbf{d}_{i+1}$
\State $\textbf{a}_i \leftarrow \textbf{a}_i - \textbf{c}_i \textbf{a}_{i+1}$
\State $\textbf{c}_i \leftarrow - \textbf{c}_i \textbf{c}_{i+1}$
\EndFor

\State $r \leftarrow 1/(1 - \textbf{a}_1 \textbf{c}_0)$
\State $\textbf{d}_0 \leftarrow r(\textbf{d}_0-\textbf{c}_0 \textbf{d}_1)$
\State $\textbf{a}_0 \leftarrow r \textbf{a}_0$
\State $\textbf{c}_0 \leftarrow -r \textbf{c}_0 \textbf{c}_1$
\State $\textbf{b} = \textbf{1}$%
\\ 
\State \textbf{Step 2: Solve reduced system of equations for boundary values}
\\
\State \textbf{Step 3: Reconstruct the solution in-place} 
\State \textbf{Input:} $\textbf{a},\textbf{c}$,$\textbf{d},\textbf{x}_0,\textbf{x}_{m-1}$
\State $\textbf{d}_0 \leftarrow \textbf{x}_0$
\State $\textbf{d}_{m-1} \leftarrow \textbf{x}_{m-1}$

\For{i=1,...,m-2}
\State $\textbf{d}_i \leftarrow \textbf{d}_i - \textbf{a}_i  \textbf{x}_0 - \textbf{c}_i \textbf{x}_{m-1}$
\EndFor
\end{algorithmic}
\end{algorithm}
A few important notes should be considered to secure parallel performance at scale when combining the pressure Poisson equation with $z$-implicit diffusion. While the same computational approach may be taken for solving both cases, the straightforward implementation of Algorithm~\ref{algo_orig_cyclic} would be far from optimal in both cases. Note that: (1) the reduced tridiagonal system for the Poisson equation is time-invariant (eq.~\eqref{proj1_Lphi}), with a problem that changes for each $(i,j)$ index, yet (2) the $z$-implicit matrix is constant for each $(i,j)$ index, but time-dependent (eq.~\eqref{proj3_impl1d}). Hence, a key optimization for the Poisson equation is to perform the transpose operations associated with the reduced system coefficients $(a',b',c')$ only once as an initialization step (recall the penultimate step in Figure~\ref{fig_FTM_2d_pencils}), as only $d'$ varies with the right-hand-side of eq.~\eqref{proj1_Lphi}. Regarding implicit $z$ diffusion, $(a',b',c')$ are time-dependent, but identical for all $(i,j)$ indexes mapped to different GPU threads to solve the equations along $z$. Thus, rather than communicating $(a',b',c')$ through MPI operations, it is much faster to have each GPU computing the global $(a',b',c')$ coefficients corresponding {to all MPI tasks aligned in the z-direction}, and locally copy the values pertaining to its own subdomain. This is done on the GPUs with unnoticeable computational overhead, and effectively avoids expensive MPI communication operations.\par
Accordingly, to efficiently handle implicit $z$ diffusion, Algorithm~\ref{algo_orig_cyclic} was re-derived in a flavor that splits the solution into an initialization with pre-computed coefficients and a runtime step. This approach is presented in Algorithm~\ref{algo_new_cyclic}. {A major advantage of this approach is that only the array $d'$ is modified in-place at runtime. This marks a large contrast to Algorithm~
\ref{algo_orig_cyclic}, which requires thread-private arrays (or memory buffers) to track intermediate changes in the arrays $(a',c')$. This change is particularly relevant when solving for implicit 1D diffusion, since only 1D arrays with precomputed $(a',b',c')$ coefficients can handle the solution process. %

From a mathematical perspective, the new algorithm is derived by analyzing the cyclic reduction process, and carefully tracking which references to the $(a',b',c')$ arrays can be replaced by either their input or output values. After making this distinction, it becomes evident that the reduction process for the array $d'$ does not depend on intermediate values for $(a',b',c')$ being over-written. Therefore, it is natural to split the process into initialization and runtime stages. Moreover, it is important to highlight that only the initial values of $(a',c')$ are used. In the DNS solver, the variables $(a',c')$ always correspond to 1D vectors, even for Poisson or Helmholtz solvers. As a result, storing the initial values of $(a,c)$ creates a negligible performance overhead. 

}

\begin{algorithm}
\caption{Alternative flavor of the cyclic reduction method proposed in this work, with a separation between initialization and runtime operations. The variables $(\textbf{A},\textbf{B},\textbf{C})$ correspond to the original coefficients of the tridiagonal equations, whereas $(\textbf{a},\textbf{b},\textbf{c})$ are the modified coefficients after cyclic reduction.}
\label{algo_new_cyclic}
\begin{algorithmic}[1]
\State \textbf{Step 1: Initialization} 
\State \textbf{Input:} $\textbf{A},\textbf{B},\textbf{C},\textbf{a},\textbf{b},\textbf{c}$
\State $\textbf{a}_1 \leftarrow {A}_1$ 
\State $\textbf{b}_1 \leftarrow \textbf{B}_1$ %

\For{i=2,...,m-1}
\State $\textbf{b}_i \leftarrow \textbf{B}_i  -{A}_i\  {C}_{i-1}/ \textbf{b}_{i-1}$
\State $\textbf{a}_i \leftarrow      -{A}_i\  \textbf{a}_{i-1}/ \textbf{b}_{i-1}$
\EndFor
\State $\textbf{c}_{m-1} \leftarrow {C}_{m-1}$
\State $\textbf{c}_{m-2} \leftarrow {C}_{m-2}$
\For{i=m-3,...,1}
\State $\textbf{a}_i \leftarrow \textbf{a}_i  -{C}_i\ \textbf{a}_{i+1}/ \textbf{b}_{i+1}$
\State $\textbf{c}_i \leftarrow      -{C}_i\ \textbf{c}_{i+1}/ \textbf{b}_{i+1}$
\EndFor
\State $\textbf{a}_0 \leftarrow {A}_0$

\State $\textbf{b}_0 \leftarrow \textbf{B}_0 -{C}_0\ \textbf{a}_{1}/ \textbf{B}_{1}$
\State $\textbf{c}_0 \leftarrow       -{C}_0\ \textbf{c}_{1}/ \textbf{B}_{1}$
\Statex \(\triangleright\) Note: To avoid GPU divisions in Steps 2 and 4, optimized implementations can store $1/\textbf{b}$ instead of $\textbf{b}$.
\\ 
\State \textbf{Step 2: Runtime reduction} 
\State \textbf{Input:} $\textbf{A},\textbf{B},\textbf{C},\textbf{a},\textbf{b},\textbf{c}$,$\textbf{d}$

\For{i=2,...,m-1}
\State $\textbf{d}_i \leftarrow \textbf{d}_i - A_i\  \textbf{d}_{i-1}/\textbf{b}_{i-1}$
\EndFor

\For{i=m-3,...,0}
\State $\textbf{d}_i \leftarrow \textbf{d}_i - C_i\  \textbf{d}_{i+1}/\textbf{b}_{i+1}$
\EndFor
\\
\State \textbf{Step 3: Solve reduced system of equations for boundary values}
\\
\State \textbf{Step 4: Reconstruct the solution in-place} 
\State \textbf{Input:} $\textbf{a},\textbf{b},\textbf{c}$,$\textbf{d},\textbf{x}_0,\textbf{x}_{m-1}$
\State $\textbf{d}_0 \leftarrow \textbf{x}_0$
\State $\textbf{d}_{m-1} \leftarrow \textbf{x}_{m-1}$

\For{i=1,...,m-2}
\State $\textbf{d}_i \leftarrow (\textbf{d}_i - \textbf{a}_i  \textbf{x}_0 - \textbf{c}_i \textbf{x}_{m-1})/\textbf{b}_{i}$
\EndFor

\end{algorithmic}
\end{algorithm}

\FloatBarrier

\FloatBarrier

\subsection{Implementation} \label{section_code_implementation}

As previously noted, the approaches we have presented are implemented in the open-source code CaNS \cite{cpu_cans_18,gpu_cans_21}. CaNS is written in Modern Fortran, and since its version \textit{2.0} offloads data and computation to GPUs using OpenACC. At the fine-grained parallelization level of the TDMA implementations, each $i,j$ index is assigned to a thread, which serially performs operations along the third domain direction. {The CPU implementation uses the \emph{2DECOMP}\&\emph{FFT} library \cite{2DECOMP} to perform transposes.} Conversely, the multi-GPU implementation uses the \emph{cuDecomp} library for pencil-distributed calculations at scale that feature transposes and halo exchanges \cite{cudecomp} is used in CaNS. The main advantage of \emph{cuDecomp} is its runtime autotuning capabilities, which allows to confidently select a well-performing combination of 2D processor grid and communication backend (with several low-level implementations of the transposing algorithm in CUDA-aware MPI, NCCL, or {NVSHMEM}). \emph{cuDecomp}'s flexibility allowed for a very straightforward implementation of the communication operations that can be visualized in Figure~\ref{fig_FTM_2d_pencils}. Indeed, the \emph{all-to-all} operations needed to communicate the boundary values for each sub-group of slices along the z-direction could be replaced by the existing transpose operations available in the \emph{cuDecomp} or \emph{2DECOMP} libraries. \par
The distributed-memory implementation of CaNS, using \emph{cuDecomp} and \emph{cuFFT}, allowed for very efficient calculations at scale {on NVIDIA-based systems}. However, in the present work we decided to benchmark our approach on the supercomputers Leonardo (NVIDIA-based) and LUMI (AMD-based). We summarize our implementation approach for the latter, which we plan to incorporate in the CaNS public repository in the near future.

{Regarding the verification of the implementation, {CaNS} has been extensively validated in the past \cite{cpu_cans_18,gpu_cans_21}. Therefore, the correctness of the current implementation can be trivially verified, since only the parallel tridiagonal solvers have been touched. An explicit comparison with respect to the output of the full-transpose method is accurate up to machine precision for the modified subroutines, which verifies the correctness of the implementation.}

\subsubsection{Many-GPU implementation on LUMI} \label{section_code_implementation_lumi}

On LUMI, the Cray Fortran compiler is readily compatible with OpenACC, and GPU-aware MPI is supported to perform data transfer among GPU devices (or GCDs). However, since LUMI has AMD cards, some work was required to port the transpose and halo exchange algorithms for LUMI. One approach would be to adjust \emph{cuDecomp} such that the NVIDIA-specific features of the library are masked out of the build workflow. In the present work, we took a different route and developed a cross-platform communication library based on Fortran and OpenACC named \emph{diezDecomp}.{ Its source code is available on GitHub under an MIT license \cite{diezdecomp}.} This implementation uses OpenACC and GPU-aware MPI to perform transpose and halo exchange operations. During transpose operations, \emph{diezDecomp} has been optimized to pack and unpack data related to different MPI ranks simultaneously {(in parallel GPU threads)}, and it supports conversions between different indexing orders (e.g., $x/y/z$ to $y/x/z$).
As further verification, the performance of the \emph{diezDecomp} library was tested in the Leonardo supercomputer, achieving nearly identical running times as the \emph{cuDecomp} library.

Finally, it is important to highlight that the current distributed Poisson/Helmholtz solver is able to work with various types of FFT libraries, such as \emph{cuFFT} \cite{cufft}, \emph{hipFFT} \cite{hipFFT}, or even in-house FFT implementations.{ In this context, our porting effort uses the \emph{hipFFT} library for simulations in AMD GPUs using the bindings provided by the \emph{hipFort} project \cite{hipFort}, whereas \emph{cuFFT} is enabled for NVIDIA GPUs via the CUDA toolkit \cite{cuda_fortran_book}. Since both FFT libraries have similar APIs, our implementation uses CPP macros to switch between libraries depending on the target platform.}

\section{Results} \label{chapter_results}
\subsection{Strong and weak scalability}

\begin{figure*}[h]
\centering
\includegraphics[width=153mm]{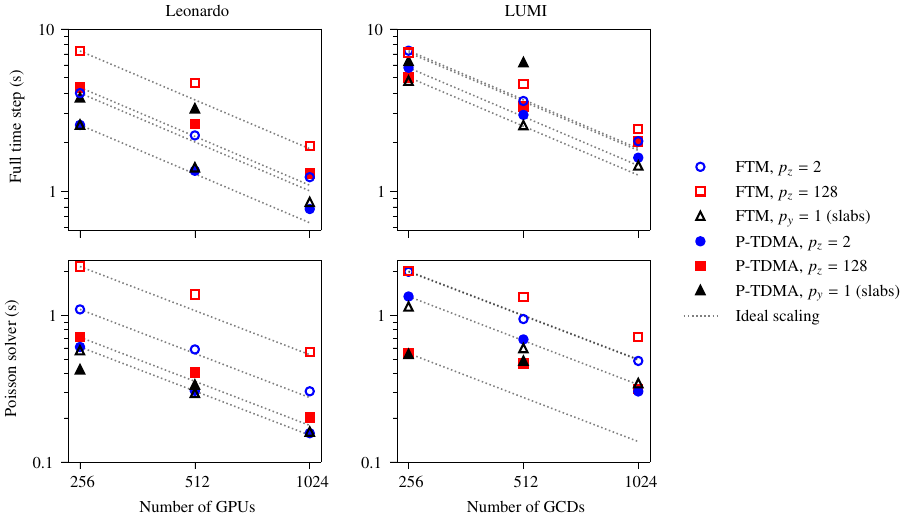}
\caption{Strong scalability chart for a wall-bounded flow with grid size $\left(N_x~\times~N_y~\times~N_z \right) = \left( 7168~\times~7168~\times~1594 \right)$, which roughly corresponds to a friction Reynolds number of $Re_{\tau}\approx5\,000$. The variable $p_z$ corresponds to the number of divisions along the $z$ direction for the pencil decomposition scheme. The abbreviation "FTM" refers to the original DNS solver, which was based on the full-transpose method. All simulations were performed in the Leonardo and LUMI supercomputers. {Please note that P-TDMA method with 1D slabs (filled black triangles) cannot be run with $1\,024$ GPUs/GCDs, due to an insufficient number of grid points in the $z$-direction: $N_z/p_z<2$.}%
}
\label{fig_strong_scalability}
\end{figure*}

\begin{figure*}[h]
\centering
\includegraphics[width=131mm]{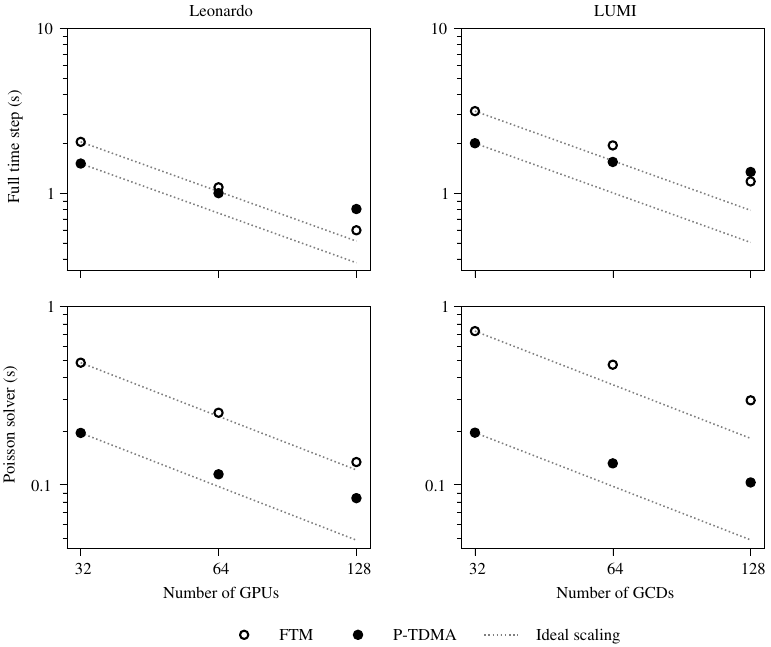}
\caption{Strong scalability chart for a wall-bounded flow with grid size $\left(N_x~\times~N_y~\times~N_z \right) = \left( 3200~\times~3456~\times~900 \right)$, which roughly corresponds to a friction Reynolds number of $Re_{\tau}\approx2,500$. All runs correspond to 1D slab configurations, as in Figure~\ref{fig_FTM_2d_pencils} without decomposing along $x$, and in Figure~\ref{fig_DTDMA_2d_pencils} without decomposing along $y$.}
\label{fig_strong_scalability_1d_slabs}
\end{figure*}

We consider a large-scale turbulent plane channel flow setup, which exercises all important steps presented in this paper, including $z$-implicit diffusion. Figure~\ref{fig_strong_scalability} presents a strong scalability test for a $N_x~\times~N_y~\times~N_z = 7168~\times~7168~\times~1594$ grid, containing approximately $80$ billion grids points. This corresponds to a channel with a domain size of $L_x~\times~L_y~\times~L_z~=~12.8~\times~6.4~\times~2$, with a friction Reynolds number $Re_{\tau}\approx5\,000$ \cite{Lee_Moser_2015}. Note that, in many GPU-resident workloads, a strong scaling test likely shows performance deterioration, as the occupancy of each GPU is being lowered \cite{cudecomp}. The tests were performed on the Leonardo and LUMI supercomputers, which again, feature NVIDIA~A100 and AMD~MI250X GPU cards, respectively. In the figure, we compare the full transpose method (FTM) to the approach with the parallelized tridiagonal matrix algorithm (P-TDMA), for both 1D (slab) decompositions (as in Figure~\ref{fig_FTM_2d_pencils} without decomposing along $x$, and in Figure~\ref{fig_DTDMA_2d_pencils} without decomposing along $y$) and for 2D pencil decomposition schemes with different levels of process decomposition along $z$: $p_z=2$ and $128$. The number of partitions along the $y$-direction ($p_y$) are set from the total number of GPUs: $n_{g} = p_y~\times~p_z$. {{These configurations are intended to show how the solver scales in the two limits where the 2D pencil decomposition has either the least number of partitions along $z$, or very high values.}} {Clearly, the P-TDMA algorithm is far more efficient than FTM when 2D pencil decompositions are considered. This is explained by Figures~\ref{fig_FTM_2d_pencils} and~\ref{fig_DTDMA_2d_pencils}, because the P-TDMA algorithm transposes a significantly smaller amount of data than the FTM method. Focusing on the performance of the Poisson solver (bottom panels of  Figure~\ref{fig_strong_scalability}), there is a consistent $1.5\times$ improvement in wall-clock time per step. While strong scaling deterioration is expected, it is interesting to note that some of the curves show reasonably mild deviations from the ideal scaling curve. 

The P-TDMA method shows great efficiency for $p_z=2$, and almost matches the best-performing 1D slab configurations in the FTM case. This is expected, as the P-TDMA method only transfers data along one small boundary in the $z$-direction when $p_z=2$, with a computational cost roughly similar to a halo exchange.%

Let us now consider the difference between the wall-clock time of the Poisson solver and the full time steps. The most important difference here is the solution of three separate systems of 1D implicit diffusion equations in the wall-clock time of the full time step. Therefore, configurations that transfer more data when solving for 1D implicit diffusion tend to have worse performance, such as the FTM approach with a 2D pencil decomposition.

The wall-time per steps for the Poisson solver with $1\,024$ GPUs/GCDs converge to a similar value for both the best FTM configurations and the P-TDMA cases with 2D pencil decompositions, since their performance is dominated by the transpose operations in the $x$-$y$ directions.{ Interestingly, one can notice a major performance gain for the P-TDMA when $p_z=128$ on LUMI, which corresponds to a special case where 50\% of the required data after the $x$-$y$ transpose is already present in the GPU, requiring only a local copy.}

In the case of the P-TDMA algorithm with 1D slabs, it can be observed in Figure~\ref{fig_strong_scalability} that the running times remain nearly constant when the number of GPUs/GCDs increases from 256 to 512. This is due to the P-TDMA method having boundaries with a fixed size of $(N_x \times N_z \times 2)$, and thus, the data transferred per GPU/GCD remains equal as the number of processes is increased.
Despite this, the P-TDMA method with 1D slabs still outperforms the FTM approach, when the number of grid points along the z-direction is very large for every 1D slab. This is illustrated in Figure~\ref{fig_strong_scalability_1d_slabs}, where a strong scalability benchmark is presented for 1D slabs for $Re_{\tau}\approx 2500$ and a grid size of $N_x~\times~N_y~\times~N_z = 3200~\times~3456~\times~900$. In this benchmark, the P-TDMA method outperforms the FTM approach with 32 GPUs/GCDs, yet its performance decreases as the number of processes grows. For cases at lower Reynolds numbers, the performance improvements can be expected to grow in favor of the P-TDMA method, since the number of grid points per GPU along $z$ increases for the 1D slabs.
}\par

\begin{figure*}[h]
\centering
\includegraphics[width=129mm]{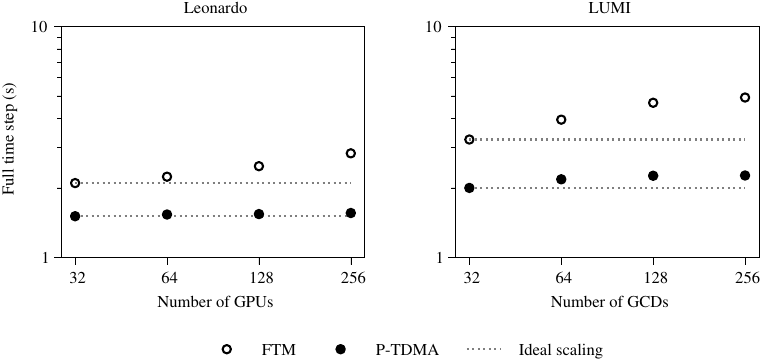}
\caption{Weak scalability analysis for DNS cases with a fixed grid size $\left(N_x~\times~N_y~\times~N_z \right) = \left( 3456~\times~3072~\times~30 \right)$ per GPU device. {For the P-TDMA method,} the pencil decomposition scheme only considers one partition along the y-direction ($p_y=1$), and a number of z-partitions ($p_z$) equal to the number of GPUs. The FTM approach considers the 1D slab configuration as in Figure~\ref{fig_FTM_2d_pencils} without decomposing along $x$.}
\label{fig_weak_scalability}
\end{figure*}

The results from the weak scaling tests are shown in Figure~\ref{fig_weak_scalability}. Weak scaling is the most important scaling indicator in large-scale GPU-resident DNS, as optimal resource usage requires maximizing GPU occupancy, which is kept fixed in these tests. We fixed the local domain sizes to about 318.5 million grid points per GPU, by considering a grid with $3456~\times~3072~\times~30$ points per GPU, which saturates the GPUs (GCDs) on Leonardo (LUMI). For the P-TDMA method, the number of partitions $p_z$ along the z-direction is equal to the number of {GPUs/GCDs}, since this corresponds to the most challenging scenario for the weak scaling analysis; for FTM, we use the optimal configuration for 1D slabs. Clearly, the performance of the P-TDMA approach is superior, not only being approximately $2\times$ faster in wall-clock time per step, but also in terms of weak scaling. An 8-fold increase in the GPUs/GCDs used results in a {3\%} (13\%) performance degradation on Leonardo (LUMI), while the full transpose method shows a major weak scaling loss of 35\% (52\%) on Leonardo (LUMI). This shows that, indeed, the proposed improvements are important for efficient wall turbulence simulations at scale.

\FloatBarrier
\subsection{Breakdown of parallel performance}

\begin{figure*}[h]
\centering
\includegraphics[width=125mm]{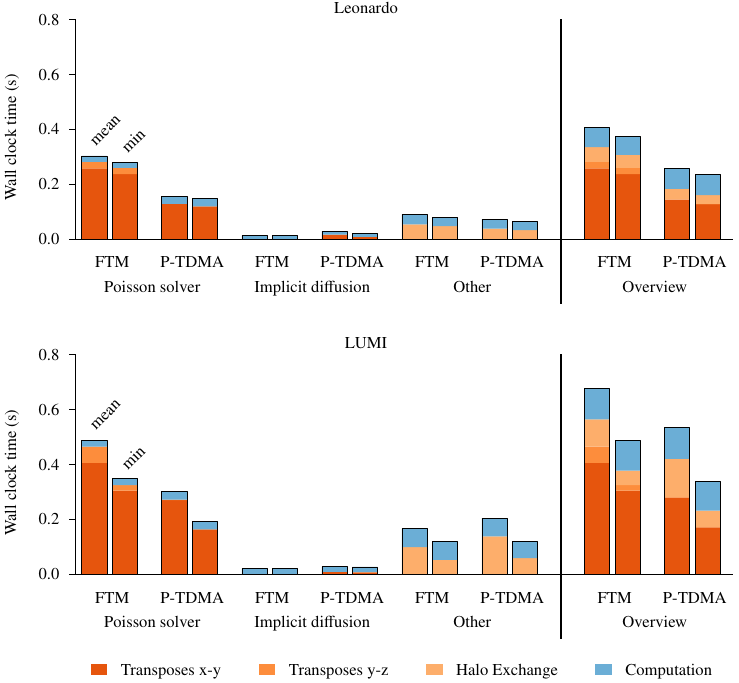}
\caption{Comparison between the GPU timer profiles for the DNS solver with the P-TDMA and FTM methods on Leonardo and LUMI. {For each profiled operation, the left bars correspond to the average wall-time per step (\emph{mean}), while the right bars are the aggregated results considering the best elapsed times per operation (\emph{min}).} The data was extracted from the DNS runs with $1\,024$ GPUs (Leonardo) or GCDs (LUMI), which are presented in the strong scalability chart (Figure~\ref{fig_strong_scalability}) for a fixed grid size $\left(N_x~\times~N_y~\times~N_z \right) = \left( 7168~\times~7168~\times~1594 \right)$. For the DNS solver with P-TDMA, the data further corresponds to the case with $\left(p_y~\times~p_z \right) = \left( 512~\times~2 \right)$ partitions along the $y$ and $z$ directions,  which has the fastest running times in the strong scalability chart. Both GPU timer profiles correspond to the averaged results for a single RK3 sub-step within the DNS solvers. Please note that the $y~\leftrightarrow~z$ transposes for the P-TDMA algorithm transfer much less data, and thus their impact on the plotted budgets is minor.}
\label{fig_gpu_profile}
\end{figure*}

{Let us now understand in more detail the results presented in the previous sections. Figure~\ref{fig_gpu_profile}, compares the workload distribution of the current approach and the previous algorithm based on the full-transpose method. The DNS cases chosen for comparison are simulations with $1\,024$ GPUs/GCDs, shown in the strong scalability chart (Figure~\ref{fig_strong_scalability}). {Consistently with the previous observations, the performance differences in the supercomputers we have tested show similar trends overall.} The 2D pencil decomposition considered for comparison is $(p_y~\times~p_z)~=~(512,~2)$, which has the lowest running times overall for the P-TDMA algorithm.  Not surprisingly, the P-TDMA approach is much faster than the full-transpose algorithm while solving for the pressure-Poisson equation. However, the P-TDMA algorithm is slower when solving for 1D implicit diffusion alongside each velocity component $(u,v,w)$. This is also expected, as again, the FTM uses an initial $z$-aligned decomposition, where the full tridiagonal problems are local to each MPI task. Both algorithms show similar overhead associated with halo exchanges, which is also expected. %
}

\subsubsection{Average timings}

When inspecting the contributions to the mean total wall-time per step in Figure~\ref{fig_gpu_profile}, it becomes clear that \emph{all-to-all} operations are the main performance bottleneck of the full-transpose method, whereas the P-TDMA approach is far less communication-intensive. Interestingly, the Poisson solver with the P-TDMA algorithm spends {81.7\% (89.6\%) of the time on Leonardo (LUMI)} performing local transposes along the $x$-$y$ plane, which are required between the Fourier-based transforms described in Section~\ref{section_poi_helm_solver}. The \emph{all-to-all} operations associated with the parallel tridiagonal solver only require {0.2\% (0.6\%)} of the computing time in Leonardo (LUMI), as this collective operation only communicates boundary values for each MPI task (recall the penultimate step in Figure~\ref{fig_FTM_2d_pencils}). Note that, while the time spent performing $x~\to~y$ transposes can be (partially) reduced by minimizing the {$p_y$ divisions along the $y$-direction (e.g., $p_y=2$)}, the overall performance degrades in Figure~\ref{fig_strong_scalability} when the number of vertical partitions $p_z$ is increased for this case. This trend is valid for both the FTM approach and the P-TDMA methods, since the cost of performing $y$-$z$ transposes increases as $p_z$ grows. Therefore, a careful trade-off must be considered. Moreover, the optimal DNS domain decomposition is not only problem-dependent but also hardware-dependent, which makes runtime tuning of the computational setup very relevant \cite{cudecomp}.

Finally, in Figure~\ref{fig_gpu_profile}, it can be noticed  that the P-TDMA approach also performs $x~\to~y$ transposes when solving for implicit 1D diffusion, with a small but noticeable communication footprint. This is not strictly necessary, since cyclic reduction can be directly performed using the initial $x$-aligned pencils (Figure~\ref{fig_DTDMA_2d_pencils}, left), and then a direct $x~\to~z$ transpose could be used to obtain $z$-aligned pencils \cite{afid_gpu}. This direct transpose is less trivial to implement and is not featured in \emph{cuDecomp} or \emph{2DECOMP}\&\emph{FFT}. Yet, since the \emph{diezDecomp} communication library supports $x~\to~z$ transposes for any desired 2D decomposition, we tested the performance gains of this direct transpose. The results of this additional benchmark are shown in \ref{section_results_xz_transp} for LUMI. {While the computational overhead of the additional $x~\to~y$ transpose is small for the cases shown in Figure~\ref{fig_gpu_profile} with the decomposition $(p_y~\times~p_z)~=~(512~\times~2)$, we identified that other pencil decompositions suffered from higher performance losses. For instance, the benchmark shows that the DNS case with $(p_y~\times~p_z)~=~(8~\times~128)$ is about $55\%$ faster during the calculation of $z$-implicit diffusion when using direct $x~\to~z$ transposes. This results in savings of about $18\%$ in the total wall-clock time per step.} 

\FloatBarrier

\subsubsection{{Best elapsed times per operation}}\label{section_top1_results}
{To better explain the differences between systems, we included in Figure~\ref{fig_gpu_profile} the aggregated best elapsed times per operation (right bars). Remarkably, the total running times for the FTM (P-TDMA) method are decreased by about 28\% (37\%) for LUMI, and by 8\% (9\%) for Leonardo. It appears that the average wall-clock time per step on LUMI is significantly suboptimal compared to that on Leonardo.}

{Significant performance differences between these machines for \emph{all-to-all} collective operations have been reported in \citet{a2a_lumi_leo_bandwidth}, where Leonardo outperformed LUMI by almost 50\% in goodput benchmarks at scale. We argue that the analysis considering the best elapsed times per operation resembles a goodput benchmark better, since these results contain the smallest levels of network delays possible. Interestingly, when inspecting the aggregated best elapsed times (\emph{min}) in Figure~\ref{fig_gpu_profile}, it can be noticed that LUMI is only 31\% (45\%) slower than Leonardo for the total running times of the FTM (P-TDMA) method, consistently with the trends in \citet{a2a_lumi_leo_bandwidth}. We verified that this trend is robust across other cases (not shown).}

\section{Conclusions}
\label{section_conclusions}

We have presented a numerical approach for GPU-based massively-parallel DNS of turbulent wall flows with one inhomogeneous direction. Using the CaNS solver as base, we extended it with a parallel tridiagonal algorithm for solving the pressure Poisson equation, and to handle implicit integration of the wall-normal diffusion term. To achieve this, we adopted a recently-proposed approach for solving distributed tridiagonal systems \cite{lazlo,pascal_tdma}, and implemented it in a pencil-decomposed framework. Allowing for two-dimensional decompositions is key, as slab-decomposed approaches are bound to breakdown for DNS at sufficiently high Reynolds number.\par
Carefully handling $z$-implicit diffusion was key to secure the improved parallel performance at scale. To this end, we have proposed a re-worked flavor of the original parallel cyclic reduction -- TDMA approach presented in \citet{lazlo}. We have shown that, by re-working the algorithm into a pre-processing and runtime step, one can {easily} solve the three linear systems per time iteration associated with this implicit discretization.\par
We have tested the different approaches at scale, using up to $1\,024$ GPUs/GCDs on the supercomputers Leonardo and LUMI. The results of the scalability tests reveal that the new distributed Poisson solver shows compelling performance gains for 2D pencil decompositions, being approximately twice faster in the LUMI and Leonardo supercomputers than the original CaNS version based on the full-transpose approach using $1\,024$ GPUs/GCDs. A detailed analysis of the GPU timer profiles reveals that the performance improvements are largely due to the reduced size of the global all-to-all transpose operations among MPI tasks. The scalability of the DNS solver was also tested in large-scale simulations of wall-bounded flows, benchmarking the performance of entire physical time steps. At scale, the new approach was found to be approximately $1.5\times$ faster in the LUMI and Leonardo supercomputers with a 2D pencil decomposition of $\left(p_y~\times~p_z \right) = \left( 512~\times~2 \right)$ while completing entire physical time steps. {The observed performance differences between the two machines were understood by inspecting the best recorded times per algorithm step, showing that LUMI runs experienced higher latency than Leonardo.} In general, we find that minimizing the number of $p_z$ partitions in 2D pencil decompositions reduces the running times for the DNS solver with either the full-transpose method or the parallel tridiagonal algorithm. This is attributed to the reduced cost of performing transposes in the $y$-$z$ directions. Additionally, we highlight that the DNS solver, coupled with the parallel tridiagonal algorithm, can be configured to work with 2D pencil decompositions achieving identical running times as the most optimized 1D slab configurations available for medium-scale systems. 

Regarding implementation, while the underlying numerical solver works on Leonardo out-of-the-box, some work was required to successfully run it on LUMI. As a by-product of the present effort, a cross-platform communication library \emph{diezDecomp} was developed for halo exchanges and any-to-any transpose operations between MPI ranks with mismatched local problem sizes. While we could have achieved the same by modifying \emph{cuDecomp}, we found some advantages in having a simpler library in Modern Fortran as an alternative with fewer dependencies.\par

Overall, this approach will enable DNS of turbulent wall flows at unprecedented scales, helping to bridge the gap between current setups that can be studied using first-principles simulations, and important applications in environmental and engineering systems.

\section{Acknowledgements}
The performance tests on the Leonardo supercomputer, based at CINECA (Italy), were enabled by the EuroHPC grant no.~\texttt{EHPC-EXT-2022E01-054}. {Access to the LUMI supercomputer was granted through the project no.~\texttt{EHPC-DEV-2024D04-039}, {and the NWO Large Compute Grant no.~\texttt{EINF-10608}}.} We also acknowledge NWO for providing access to the Snellius supercomputer, based at SURF (Netherlands). We thank Josh Romero, Massimiliano Fatica, and Sergio Pirozzoli for the insightful discussions. Finally, we would also like to thank the two anonymous Reviewers for their careful review, which contributed to the improvement of the manuscript.
\appendix
\section{GPU implementation of Fourier-based transform}
\label{appendix_fourier}
While a real-to-complex Fourier transform of a signal $\mathbf{x}$ with $n$ numbers has $n/2+1$ complex numbers, the imaginary parts of the first and last elements {(for even $n$)} are zero. Let us consider the output of the real-to-complex Fourier transform of 
\begin{equation}
\label{eq_ApB_xreal}
\textbf{x} =
\begin{bmatrix*}[l]
  x_0 & x_1 &  \dotsc & x_{n-1}
\end{bmatrix*} ,
\end{equation}
given by
\begin{equation}
\label{eq_ApB_ximag}
\bm{\tilde{x}} =
\begin{bmatrix*}[l]
\tilde{x}_0^r & \tilde{x}_0^i & \tilde{x}_1^r & \tilde{x}_1^i & \dotsc & \tilde{x}_{\lfloor n/2 + 1 \rfloor}^r & \tilde{x}_{\lfloor n/2 + 1 \rfloor}^i
\end{bmatrix*};
\end{equation}
$\bm{\tilde{x}}$ has $\lfloor n/2 +1 \rfloor$ elements, with $\lfloor \rfloor$ denoting the integer floor operation. Since each complex number is represented by two real ones, $\bm{\tilde{x}}$ is represented by $(2\ \lfloor n/2 + 1\rfloor)\ $ real numbers, with $\tilde{x}_0^i=0$, and $\tilde{x}_{\lfloor n/2 + 1 \rfloor}^i = 0$ for even $n$. Hence, the real-to-complex transform can be uniquely represented by a set of $n$ numbers. This property is explored in several FFT packages (e.g., {\emph{FFTPACK}}, and the \emph{half-complex} format of FFTW used in the CaNS code for CPU-based runs). Unfortunately, popular GPU-based FFT libraries like \emph{cuFFT}, \emph{hipFFT}, or \emph{MKL} do not support this format \cite{cufft,hipFFT,MKL}.

Representing the output of the real-to-complex transforms in arrays of size $n$ is desirable, as it allows us to handle the output of a real-to-complex transform in the same manner as a real-to-real transform, greatly simplifying the implementation of different transform types in the Poisson solver. Hence, since the first GPU version of the CaNS code \cite{gpu_cans_21}, $\bm{\tilde{x}}$ is packed in the following format:

\begin{equation}
\label{eq_ApB_ximag2}
  \bm{\tilde{x}'} =
  \begin{cases}
  \begin{tikzpicture}[baseline={([yshift=-.5ex]current bounding box.center)},
  start chain = going right, node distance = 0pt,
MyStyle/.style={draw, minimum width=3.8em, minimum height=2em, outer sep=0pt, on chain,line width=0.25pt}]
\node [MyStyle] (1) {$\tilde{x}_0^r$};
\node [MyStyle] (2) {$\tilde{x}_0^i$};
\node [MyStyle] (3) {$\dotsc$};
\node [MyStyle,fill=tikzlightgray] (4) {$\tilde{x}_{\lfloor n/2 + 1 \rfloor }^r$};
\node [MyStyle,fill=tikzlightgray] (5) {$\tilde{x}_{\lfloor n/2 + 1 \rfloor }^i$};
\draw[->] (4) --++(0,-1.5em) -| (2);
\end{tikzpicture} &  \text{if $n$ is even,} \\
  \begin{tikzpicture}[outer sep = 5em,baseline={([yshift=-.5ex]current bounding box.center)},
  start chain = going right, node distance = 0pt,
MyStyle/.style={draw, minimum width=3.8em, minimum height=2em, outer sep=0pt, on chain,line width=0.25pt}]
\node [MyStyle] (1) {$\tilde{x}_0^r$};
\node [MyStyle] (2) {$\tilde{x}_0^i$};
\node [MyStyle] (3) {$\dotsc$};
\node [MyStyle] (4) {$\tilde{x}_{\lfloor n/2 + 1 \rfloor }^r$};
\node [MyStyle,fill=tikzlightgray] (5) {$\tilde{x}_{\lfloor n/2 + 1 \rfloor }^i$};
\draw[->] (5) --++(0,-1.5em) -| (2);
\end{tikzpicture} &  \text{if $n$ is odd.} \\
\end{cases}
\end{equation}
It is easily seen that both cases have $n$ real elements. This operation has $\mathcal{O}(1)$ time complexity, while re-arranging the signal such as $[\tilde{x}_0^r,\ \dotsc,\ \tilde{x}_1^i,\ \dotsc]$, would have $\mathcal{O}(n)$ complexity. With this cheaper re-arrangement of the arrays, the Fourier eigenvalues $\lambda_i$ and $\lambda_j$ in eq.~\eqref{fijk_eq} must be consistently re-ordered to comply with this format. This is an inexpensive operation that is performed during the initialization of the Poisson/Helmholtz solver.

Finally, the reciprocate unpacking operations are done for performing the inverse complex-to-real transform to have an input array with $\lfloor n/2 +1 \rfloor$ elements, resulting in an output signal with $n$-elements in the correct order.

\FloatBarrier

\section{Performance gains from direct $x~\to~z$ transposes with implicit 1D diffusion}
\label{section_results_xz_transp}
\begin{figure*}[h]
\centering
\includegraphics[width=134mm]{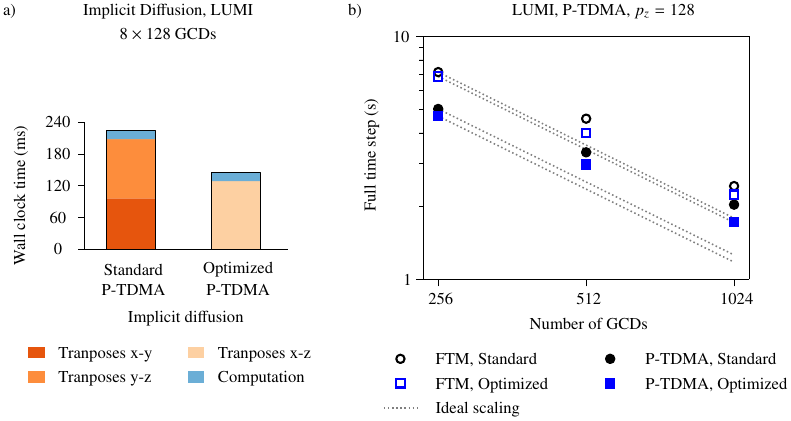}
\caption{Comparison between the standard P-TDMA approach and the optimized version using direct $x~\to~z$  transpose operations for implicit 1D diffusion (subplot a) using $8 \times 128$ GCDs and a grid size of $\left(N_x~\times~N_y~\times~N_z \right) = \left( 7168~\times~7168~\times~1594 \right)$. The subplot (b) corresponds to a strong scalability test for the entire DNS solver using the same grid size, but a different number of GCDs in the LUMI supercomputer.} %
\label{fig_collage_txz}
\end{figure*}

{When the GPU profiling results from Figure~\ref{fig_gpu_profile} are analyzed, it can be noticed that the implicit 1D diffusion solver performs two consecutive transposes in the $x$-$y$ and $y$-$z$ directions, which will be denoted as $x~\to~y~\to~z$ in this section. This is suboptimal. Ideally, the cyclic reduction process should be performed using the original $x$-aligned pencils for the velocity components $u/v/w$, and then direct $x~\to~z$ transposes should be used before solving the reduced systems of tridiagonal equations. %

To better understand the benefits of performing a direct $x~\to~z$ transpose instead of two consecutive transposes, we used the \emph{diezDecomp} library created for the LUMI porting effort. The implementation intersects the $x/y/z$ bounds of different MPI tasks, with no strong restrictions, and thus it is trivial to implement any variant of $x~\to~z$ transpose. This allowed for an implementation of this more complex communication operation with minimal changes in the DNS code.

{While avoiding $x~\to~y~\to~z$ transposes has a small impact in the GPU profiling results shown in Figure~\ref{fig_gpu_profile}, we identified other DNS cases where the impact of this transpose sequence was much higher. For instance, the DNS cases with $(p_y~\times~p_z)~=~(8~\times~128)$ have a much higher MPI workload for the parallel tridiagonal solver (due to the reduced size of $p_y$), and thus they benefit more from removing $x~\to~y~\to~z$ transposes.} In Figure~\ref{fig_collage_txz}, the results of the scalability tests using $x~\to~z$ transposes are presented for a 2D pencil decomposition with $p_z = 128$ vertical partitions. The configuration $p_z=128$ was chosen, since its parallel tridiagonal solver works with larger arrays and the unnecessary $x~\to~y$ transposes have a significant impact in the results previously shown in Figure~\ref{fig_strong_scalability}. In the subplot (a), it can be seen that the running times for the implicit 1D diffusion solver are 55\% slower when successive $x$-$y$ and $y$-$z$ transposes are used. The scalability chart at the right reveals that the system with $x~\to~z$ transposes is more efficient in large-scale simulations, reducing the running times of the entire DNS solver by 18\% for the P-TDMA algorithm with $1\,024$ GCDs.
}

\FloatBarrier
\bibliography{scalar}

\end{document}